\documentclass[12pt]{article}
\usepackage{amsfonts,amsmath}

\renewcommand{\theequation}{\thesection.\arabic{equation}}
\renewcommand{\thesubsection}{\arabic{section}.\arabic{subsection}}
\makeatletter \@addtoreset{equation}{section} \makeatother

\def\vac{|0\rangle\langle 0|}
\def\Fock{|C(b|X)\rangle}
\def\al{\alpha}

\newcommand{\be}{\begin{equation}}
\newcommand{\ee}{\end{equation}}
\newcommand{\bee}{\begin{eqnarray}}
\newcommand{\beee}{\begin{array}}
\newcommand{\eee}{\end{eqnarray}}
\newcommand{\eeee}{\end{array}}
\newcommand{\ga}{\alpha}
\newcommand{\gb}{\beta}
\newcommand{\gga}{\gamma}

\newcommand{\gd}{\delta}
\newcommand{\gl}{\lambda}
\newcommand{\gk}{\kappa}
\newcommand{\gep}{\epsilon}

\newcommand{\go}{\omega}

\newcommand{\nn}{\nonumber}

\newcommand{\p}{\partial}

\newcommand{\f}{\frac}

\begin{document}

\begin{flushright}
FIAN/TD/12-06
\end{flushright}

\vspace{0.5cm}

\begin{center}
{\large\bf BTZ Black Hole as  Solution of\\

\vspace{0.2 cm}
 $3d$ Higher Spin Gauge
Theory}

\vspace{1 cm}

{\bf V.E.~Didenko$^{1,2}$, A.S.~Matveev$^{1,2}$ and M.A.~Vasiliev$^{1}$}\\
\vspace{0.5 cm} {\it
$^1$ I.E. Tamm Department of Theoretical Physics, Lebedev Physical Institute,\\
Leninsky prospect 53, 119991, Moscow, Russia \\
\vspace{0.2 cm} $^2$ Museo Storico della Fisica e Centro Studi e
Ricerche
"Enrico Fermi", Rome, Italy} \\

\vspace{0.6 cm}
didenko@lpi.ru, matveev@lpi.ru, vasiliev@lpi.ru \\
\end{center}

\vspace{0.4 cm}

\begin{abstract}
\noindent BTZ black hole is interpreted as exact solution of $3d$
higher spin gauge theory. Solutions for free massless fields in
BTZ black hole background are constructed with the help of the
star-product algebra formalism underlying the formulation of $3d$
higher spin theory.  It is shown that a part of higher spin symmetries
remains unbroken for special values of the BTZ parameters.
\end{abstract}

\section{Introduction}

An important difference of (2+1) gravity
\cite{1963,1966,3d1,3d2,3d3} from higher dimensional gravitational
theories is that the vacuum theory is topological, describing no
local degrees of freedom. It was shown in \cite{Townsend,Wit} that
(2+1) gravity is equivalent to the Chern-Simons gauge theory of
$SL(2|\mathbb{R})\times SL(2|\mathbb{R})$ in which the gauge
potential describes dreibein and Lorentz connection. Among other
things, on mass shell, this formulation allows one to treat
diffeomorphisms of general relativity as gauge transformations,
that essentially simplifies the quantum analysis \cite{Carlip}.

In three dimensions, the Riemann tensor is fully represented by
the Ricci tensor. As a result, $R_{mn}=0$ implies $R_{mnpq}=0$,
i.e., any vacuum solution is locally Minkowski. Analogously, any
vacuum solution of the Einstein equations with negative
cosmological term is locally $AdS_3$.

BTZ black hole solution in {\it AdS$_3$} was  discovered in
\cite{BTZ}. The  ``No Black Hole Theorem" \cite{nobh}  states that
no black hole type solution with (non-zero) horizons in 2+1
dimensions exists unless negative cosmological constant is
introduced.

The BTZ black hole is in many respects analogous to the
four-dimensional Kerr black hole, thus providing a useful model for
the study of black hole physics. The important difference is however
that the BTZ black hole has  no curvature singularity \cite{BHTZ}.
The black hole type geodesics behaviour results instead from the
topological peculiarity of the BTZ solution which is locally
isomorphic to $AdS_{3}$. As shown in \cite{BHTZ}, the BTZ solution
can be  obtained via factorization of $AdS_{3}$ over a discrete
symmetry group.

Since BTZ solution  has zero $o(2,2)$ curvature, it is also the
exact solution of  the nonlinear $3d$ higher spin (HS) gauge
theory \cite{3HS,PV} which, for the case of vanishing matter fields,
amounts to the Chern-Simons theory for the $3d$ HS algebra that
contains $o(2,2)\sim sp(2)\oplus sp(2)$ as a subalgebra. Until now
a very few  exact solutions in the nonlinear HS gauge theory are
known apart from pure $AdS_d$.
One is the Lorentz invariant $3d$ solution found in \cite{PV} and
its generalization to the $4d$ HS theory obtained in \cite{Gol}.
Unfortunately,
 the physical interpretation of these solutions is still
lacking although they are likely to play a fundamental role in the
HS theory as the basis solutions for the application of the
integrating flow machinery \cite{PV}. Recently,  new exact solutions
have been found by Sezgin and Sundell \cite{SS},
 which may receive some interpretation in the AdS/CFT context.

The investigation of black hole solutions in higher-dimensional HS
gauge theories is, of course, of primary importance. The main
motivation for this work is that, although being very simple,  the
study of the BTZ black hole in the $3d$
 gauge theory can be useful for the study of less trivial
Schwarzschild-Kerr-type solutions at least in two respects.
Firstly, we learn how the HS star-product machinery applies to the
black hole physics. This is the aim of this paper. Secondly,
pretty much as $4d$ Minkowski space-time is a slice of the flat
ten-dimensional space-time with  matrix coordinates
\cite{F,BLS,BHS} $X^{AB}=X^{BA}$ ($A,B$ is the $4d$ Majorana
spinor index), it is tempting to speculate that the $4d$ Kerr
black hole can be interpreted as a slice of a BTZ-like  solution
associated with  the group manifold $Sp(4)$ which represents the
$AdS$-like geometry
 in this framework \cite{BLPS,BHS,DV,PST}.
If so, the BTZ-like zero curvature  solutions may  shed some light
on the study of the usual black hole physics from a more general
perspective of higher-dimensional generalized spaces with matrix
coordinates.

The modest aim of this letter is to demonstrate how the methods of
HS gauge theory can be applied to reproduce the known results of the
BTZ black hole physics. Namely, using the oscillator realization of the
$AdS_3$ isometry algebra $o(2,2)\sim sp(2)\oplus sp(2)$ we find the
gauge function of the BTZ solution in terms of $Sp(2)$ group and
then solve free massless field equations in the BTZ background in
terms of the Fock module \cite{SV} to show how
solutions for massless scalar and spinor fields
\cite{scalarBTZ1, scalarBTZ2, BTZspinor, BSS} are obtained in our
approach.

The layout of the rest of the paper is as follows. In Section
\ref{BTZsec} we summarize basic facts on BTZ black hole metric, its
symmetries and factorization procedure. In Section \ref{oscill} we
recall the oscillator realization of $AdS_3$ algebra. In Section
\ref{flatconect} coordinate-free description of BTZ black hole as a
flat connection is given. BTZ gauge function is represented in
Section \ref{gaugefunc}. In Section \ref{Fockmod} we review
dynamical $Sp(4)$ covariant equations and Fock module formulation of
\cite{DV,SV}. Star-product realization of Killing vectors on the
Fock module is obtained in Section \ref{killings}. In Section
\ref{solutions} we find explicit solutions for dynamical fields in
BTZ background using unfolded dynamics approach. In Section
\ref{extreme} we discuss briefly the case of extremal BTZ black
hole.  Finally, in Section \ref{Symmetry} we explore  symmetries of
massless fields in BTZ black hole background.
 Some useful formulae and intermediate calculations
are given in two Appendices.

\section{BTZ black hole}\label{BTZsec}

In this section we briefly recall some properties of BTZ black hole.
For more detail  we refer the reader to the review \cite{BHTZ}.

The $3d$ Einstein-Hilbert action with negative cosmological constant
$\Lambda=-\gl^2$
\be
S=\f{1}{2\pi}\int\sqrt{-g}(R+2\gl^2)dt d^2 x
\ee
gives Einstein equations
\be\label{Eins}
R_{mn}-\f{1}{2}Rg_{mn}=\gl^2 g_{mn}\,.
\ee
In three dimensions this implies
\be
R_{mnpq}=-\gl^2 (g_{mp}g_{nq}-g_{np}g_{mq}).
\ee
This means that, being a vacuum solution, $3d$ black hole is locally
equivalent to $AdS_{3}$. In \cite{BTZ} it was shown that the metric
\be\label{BTZ}
ds^2=(-M+\gl^2 r^2+\f{J^2}{4r^2})dt^2-(-M+\gl^2
r^2+\f{J^2}{4r^2})^{-1}dr^2-r^2(d\phi-\f{J}{2r^2}dt)^2,
\ee
where $\phi\in[0,2\pi]$, solves \eqref{Eins} and describes a
rotating black hole with dimensionless mass\footnote{Units are
chosen so that $G$ = 1/8.} $M$ and angular momentum $J$. It has the
inner and outer horizons
\be
r^2_{\pm}=\f{M}{2\gl^2}\left(1\pm\sqrt{1-\f{J^2\gl^2}{M^2}}\right)\,.
\ee
The  ergosphere (i.e., the  $g_{00}=0$ surface of infinite redshift)
has
$
r_{erg}=\f{1}{\gl}M^{1/2}.
$

Note that $r_{\pm}$ are complex for $|J| > M/\lambda$, in which case
the horizons disappear and the metric has naked singularity at
$r=0$. Formally, one can take negative $M$ in the metric \eqref{BTZ}
when $J=0$. But, except for $M=-1$ corresponding to the $AdS_{3}$
space-time, this leads to the naked conical singularity at $r=0$
\cite{BHTZ}, which is easily seen by the rescaling of
the radial variable $r \rightarrow \sqrt{-M} r$. The case of $M=0$
and $J=0$ corresponds to ``massless'' black hole that does not
reproduce the $AdS$ space-time (unlike the $4d$ case). So, we demand
\be
M>0\,,\qquad |J|\leq M/\gl.
\ee
The limiting case of $|J|=M/\gl$ corresponds to the extremal black
hole with $r_{+}=r_{-}$.

The $AdS_3$ space-time is a quadric in the four-dimensional
pseudo-Euclidian space with the metric $\eta=diag(+ + - -)$
\be\label{flat}
ds^2=du^2+dv^2-dx^2-dy^2,
\ee
\be\label{emb}
u^2+v^2-x^2-y^2=\gl^{-2}.
\ee
The metric of generic BTZ black hole \eqref{BTZ} for $r>r_{+}$ is
conveniently parameterized by
\begin{eqnarray} \label{btzcoord}
  u &=& \sqrt{A(r)} \cosh (\tilde{\phi}(t,\phi)), \nn \\
  v &=& \sqrt{B(r)} \sinh (\tilde{t}(t,\phi)),  \nn\\
  x &=& \sqrt{A(r)} \sinh (\tilde{\phi}(t,\phi)),  \nn\\
  y &=& \sqrt{B(r)} \cosh (\tilde{t}(t,\phi)),
\end{eqnarray}
where
\be\label{ABdef}
A(r)=\frac{1}{\lambda^2} \left( \frac{r^2-r^2_-}{r^2_+-r^2_-}
\right), \quad B(r)=\frac{1}{\lambda^2} \left(
\frac{r^2-r^2_+}{r^2_+-r^2_-} \right),
\ee
\be\label{tphitilde}
\tilde{t}=\lambda^2 r_+ t - \lambda r_- \phi, \quad
\tilde{\phi}=-\lambda^2 r_- t + \lambda r_+ \phi.
\ee
In this paper,  we will use the embedding relations \eqref{btzcoord}
with $r>r_{+}$ for the case of generic black hole. (For more detail
on other patches with $r\leq r_{+}$ as well as on the cases of
extremal and vacuum black holes we refer the reader to \cite{BHTZ}).

The properties of BTZ black hole are heavily based on its group
origin. Indeed, one can combine $(u, v, x, y)$ into a $2 \times 2$
matrix $S_0 \in SL(2|\mathbb{R})$
\begin{equation} \label{S-matrix}
    S_0 = \lambda\left(
      \begin{array}{cc}
        u+x & v-y \\
        -v-y & u-x \\
      \end{array}
    \right), \quad \det(S_0)=1.
\end{equation}
As shown in \cite{BHTZ}, the BTZ solution results from
the $SL(2|\mathbb{R})$ group manifold, via
factorization over a discrete subgroup by the identification
\be\label{fact}
{S_0} \sim \rho^{+}{S_0}\rho^{-}\,,\quad \rho^{\pm}=\left(
\begin{array}{cc}
e^{\pi\gl (r_{+}\pm r_{-})} & 0\\
0 & e^{-\pi\gl (r_{+}\pm r_{-})}
\end{array} \right),
\ee
which makes cyclic the variable $\phi$  in the metric \eqref{BTZ}.

The isometries of $AdS_3$ are represented as elements of the group
$SL(2|\mathbb{R})_L \times SL(2|\mathbb{R})_R/\mathbb{Z}_2 \sim
SO(2, 2)$ acting on group elements by left and right
multiplication ${S_0}\rightarrow P_L\, {S_0}\, P_R$ with the
identification $(P_L,P_R)\sim(-P_L,-P_R)$. In accordance with
\eqref{flat}, the $AdS_3$ space-time is invariant under the
$SO(2,2)$ transformations generated by
\begin{equation} \label{momentum}
J_{ab}=X_b\frac{\partial}{\partial X^a}-X_a\frac{\partial}{\partial
X^b}\,,\quad \textnormal{where}\quad X^a=(u,v,x,y).
\end{equation}
According to \cite{BHTZ}, the isometry algebra of a general BTZ
metric (\ref{BTZ}) is generated by the vector fields $\f{\p}{\p t}$
and $\f{\p}{\p \phi}$. In the case $r^2_+-r^2_->0$, the Killing
vector responsible for the identification \eqref{fact} is
\begin{equation}\label{phikil}
    \frac{\partial}{\partial \phi} = -\lambda r_+ J_{12} + \lambda r_-
    J_{03},
\end{equation}
whereas the time translation generator is
\begin{equation}\label{tkil}
    \frac{\partial}{\partial t} = \lambda^2 r_- J_{12} - \lambda^2 r_+ J_{03}.
\end{equation}
Note that, as shown in \cite{BHTZ}, among six Killing vectors of
$AdS_{3}$ only  \eqref{phikil} and \eqref{tkil} remain
globally defined  upon the identification \eqref{fact}.

\section{Oscillator realization of $o(2,2)$}\label{oscill}

Let us describe the oscillator realization of the algebra $o(2,2)$
which will be  particularly useful for our analysis. The isometry
algebra of $AdS_3$ is $o(2,2)\sim sp(2)\oplus sp(2)$. It is spanned
by the diagonal $sp(2)$ Lorentz generators $L_{\al\gb}=L_{\gb\al}$
and $AdS_3$ translations $P_{\al\gb}=P_{\gb\al}$ $(\ga,\gb,\ldots =
1,2)$.   The commutation relations are
\[
[L_{\al\gb},L_{\gga\gd}]=\f12(\gep_{\gb\gga}L_{\al\gd}+\gep_{\gb\gd}L_{\al\gga}
+\gep_{\al\gga}L_{\gb\gd}+\gep_{\al\gd}L_{\gb\gga})\,,
\]
\be
[P_{\al\gb},
P_{\gga\gd}]=2\gl^2(\gep_{\gb\gga}L_{\al\gd}+\gep_{\gb\gd}L_{\al\gga}
+\gep_{\al\gga}L_{\gb\gd}+\gep_{\al\gd}L_{\gb\gga})\,,
\ee
\[
[L_{\al\gb},
P_{\gga\gd}]=\f12(\gep_{\gb\gga}P_{\al\gd}+\gep_{\gb\gd}P_{\al\gga}
+\gep_{\al\gga}P_{\gb\gd}+\gep_{\al\gd}P_{\gb\gga})\,,
\]
where
\[ \gep_{\al\gb}=\left(
\begin{array}{cc}
0 & 1\\
-1 & 0
\end{array} \right)
\]
is the antisymmetric $sp(2)$ invariant form\footnote{Spinor indices
are raised and lowered according to the rules
$A_{\al}=A^{\gb}\gep_{\gb\al}$, $A^{\al}=\gep^{\al\gb}A_{\gb}$.}.

Let $\hat{a}_{\alpha}$ and
$\hat{b}^{\alpha}$ be oscillators with the commutation relations
\be
\label{osci}
[\hat{a}_{\alpha},\hat{b}^{\beta}]=\delta_{\alpha}{}^{\beta}
\,,\qquad [\hat{a}_{\alpha},\hat{a}_{\beta}]=0 \,,\qquad
[\hat{b}^{\alpha},\hat{b}^{\beta}]=0\,.
\ee
The generators of $sp(2)\oplus sp(2)$ admit the standard oscillator
realization \cite{Bars}
\be
\hat{L}_{\alpha}{}^{\gb}
=\frac{1}{2}\{\hat{a}_{\alpha},\hat{b}^{\beta}\}-
\frac{1}{4}\{\hat{a}_{\gga},\hat{b}^{\gga}\}\gd_{\al}{}^{\gb}
\,,\qquad \hat{P}_{\alpha\beta}=\hat{a}_{\alpha}\hat{a}_{\beta}
+\gl^2\hat{b}_{\alpha}\hat{b}_{\beta}\,.
\ee
Instead of working with operators, it is more convenient to use the
star-product operation in the algebra of polynomials of commuting
variables $a_{\alpha}$ and $b^{\alpha}$
\be\label{old}
(f\star g)(a,b)=\frac{1}{\pi^{4}}\int
f(a+u,b+t)g(a+s,b+v)e^{2(s_{\alpha}
t^{\alpha}-u_{\alpha}v^{\alpha})} \,
d^{2}u\,d^{2}t\,d^{2}s\,d^{2}v\,.
\ee
Equivalently,
\[
(f\star g)(a,b)=f(a,b)\,e^{\f12\Big(\overleftarrow{\f{\p}{\p
a_{\al}}}\overrightarrow{\f{\p}{\p b^{\al}}}-\overleftarrow{\f{\p}{\p
b^{\al}}}\overrightarrow{\f{\p}{\p a_{\al}}}\Big)}\,g(a,b)\,.
\]
The star-product defined this way (often called Moyal product)
describes the associative  product of symmetrized (i.e., Weyl
ordered) polynomials of oscillators in terms of symbols of
operators. The integral is normalized so that $1$ is the unit
element of the algebra. In particular,
$$
1 \star 1 = \frac{1}{\pi^{4}}\int e^{2(s_{\alpha}
t^{\alpha}-u_{\alpha}v^{\alpha})} \,
d^{2}u\,d^{2}t\,d^{2}s\,d^{2}v=1 .
$$
From \eqref{old} it follows that
\begin{eqnarray} \label{stprprop}
\nonumber  a_{\alpha} \star f(a,b)&=& a_{\alpha} f(a,b) + \frac12\frac{\partial}{\partial b^{\alpha}} f(a,b), \\
\nonumber   b_{\alpha} \star f(a,b)&=& b_{\alpha} f(a,b) +
\frac12\frac{\partial}{\partial a^{\alpha}} f(a,b)\,.
\end{eqnarray}
In particular, the defining relations of the associative
star-product algebra are
\be
[a_{\alpha},b^{\beta}]_{\star}= \delta_{\alpha}{}^{\beta},\qquad
{[}a_{\alpha},a_{\beta}]_{\star}=0,\qquad
{[}b^{\alpha},b^{\beta}]_{\star}=0,
\ee
where $[a,b]_{\star} = a\star b - b\star a$. The star-product
realization of the $o(2,2)$ generators is
\be \label{plgens}
L_{\alpha\beta}=\frac12
(a_{\alpha}b_{\beta}+a_{\beta}b_{\alpha})\,,\qquad
P_{\alpha\beta}=a_{\alpha}a_{\beta}+\gl^2 \, b_{\alpha}b_{\beta} \,.
\ee
For convenience, from now on we set the $AdS_3$ radius equal to
unity ($\gl$=1).

\section{BTZ black hole as flat connection}\label{flatconect}

Since  BTZ black hole  is locally equivalent to $AdS_3$ it can be
described by a flat connection  of  $sp(2)\oplus sp(2)$. Indeed, let
$w_0$ be a  $sp(2)\oplus sp(2)$ valued 1-form
\be\label{connect} w_0(a,b|X)=\frac12
\omega^{\alpha\beta}(X)L_{\alpha\beta} + \frac14
h^{\alpha\beta}(X)P_{\alpha\beta}\,, \ee where $P_{\al\gb}$ and
$L_{\al\gb}$ are the $AdS_3$ generators \eqref{plgens} while
$\go_{\al\gb}(X)$ and $h_{\al\gb}(X)$ are 1-forms. Then the
zero-curvature condition
\be\label{zero} R=dw_0-w_0\star\wedge w_0=0 \ee is equivalent to
the equations \be\label{einstein1} d\omega_{\alpha\beta}+ \frac12
\omega_{\alpha}{}^{\gamma}\wedge\omega_{\gamma\beta} +\frac12
h_{\alpha}{}^{\gamma}\wedge h_{\beta\gamma}=0, \ee
\be\label{einstein2} dh_{\alpha\beta} +\frac12
\omega_{\alpha}{}^{\gamma}\wedge h_{\gamma\beta} +\frac12
\omega_{\beta}{}^{\gamma}\wedge h_{\alpha\gamma}=0.
\end{equation}
Identifying $\go_{\al\gb}$ with Lorentz connection and $h_{\al\gb}$
with dreibein, \eqref{einstein2} gives the zero torsion condition
while \eqref{einstein1} implies local $AdS_3$ geometry.

The equation \eqref{zero} is invariant under the gauge
transformations
\be\label{gauge}
\delta w_0=d\epsilon-[w_0 ,\epsilon]_{\star}\,,
\ee
where $\epsilon(a,b|X)$ is an arbitrary infinitesimal gauge
parameter. Any fixed vacuum solution $w_{0}$ of the equation
\eqref{zero} breaks the local symmetry to its stability subalgebra
with the infinitesimal parameters $\epsilon_{0}(a,b|X)$ satisfying
the equation
\be\label{eps}
d\epsilon_{0}-[w_{0},\epsilon_{0}]_{\star}=0.
\ee
Consistency of this equation is guaranteed by \eqref{zero}. Its
generic solution has at most six independent parameters,
 the global symmetry parameters. How many of these survive in
a locally $AdS_3$ geometry depends on its global properties (i.e.,
boundary conditions). The true $AdS_3$ space-time has all six
symmetries which are $o(2,2)$ motions of $AdS_3$. For the
generic BTZ black hole solution only two of the six parameters
survive.

Locally, the general form of the  dreibein and Lorentz connection of
$sp(2)\oplus sp(2)$ algebra that satisfy  \eqref{einstein1} and
\eqref{einstein2} is
\begin{align}\label{gcalfi2}
h_{\alpha\beta} &=(W_1^{-1})_\alpha{}^\gamma d(W_1)_{\gamma\beta}
-(W_2)_\alpha{}^\gamma d(W_2^{-1})_{\gamma\beta},\\
\omega_{\alpha\beta} &=(W_1^{-1})_\alpha{}^\gamma
d(W_1)_{\gamma\beta} +(W_2)_\alpha{}^\gamma
d(W_2^{-1})_{\gamma\beta},\label{gcalfi1}
\end{align}
where $W_{1,2\,\al}{}^{\gb}(X) \in Sp(2)$, i.e.,
\begin{equation}\label{sp2prop}
    (W_{1,2}^{-1})_{\alpha\beta}=-(W_{1,2})_{\beta\alpha}.
\end{equation}
From \eqref{gcalfi2} it follows that the metric is
\begin{equation}\label{hh}
 ds^2=\f12h_{\alpha\beta}h^{\alpha\beta}=\f12dS_{\alpha\beta}dS^{\alpha\beta}\,,
\end{equation}
where
\be\label{Sw1w2}
S_{\al\gb}=(W_1)_{\al}{}^{\gga}(W_2)_{\gga\gb}.
\ee
Thus, any locally $AdS_3$  metric is determined  by a $Sp(2)$
matrix field $S_{\al\gb}(X)$. (Note that, generally,
$S_{\al\gb}\neq S_{\gb\al}$.) To obtain the BTZ metric \eqref{BTZ}
one can use the matrix $S_0$ \eqref{S-matrix} with the parametrization
\eqref{btzcoord}.

A class of
the dreibeins \eqref{gcalfi2} and Lorentz connections
\eqref{gcalfi1}, that are well-defined with respect to the identification
$\phi\to\phi+2\pi$, can be found using the following decomposition of
the matrix $S_0$
\eqref{S-matrix}:
\be
\label{san}
S_{0\,\al}{}^{\gb}=(K_{+}U_{r}K_{-})_{\al}{}^{\gb}\,
\ee
with the $Sp(2)$ matrices $K_{\pm}$ and $U_{r}$ of the form
\begin{equation} \label{KUK}
K_{\pm} = \left(
\begin{array}{cc}
e^{\f12(\tilde{\phi}\mp\tilde{t})} & 0 \\
0 & e^{-\f12(\tilde{\phi}\mp\tilde{t})} \\
\end{array}
\right)\,,\quad U_{r} = \left(
\begin{array}{cc}
\sqrt{A} & -\sqrt{B} \\
-\sqrt{B} & \sqrt{A} \\
\end{array}
\right)\,.
\end{equation}
Note, that $K_{\pm}$ belong to the Abelian  BTZ Killing subgroup
of $Sp(2)\times Sp(2)$.

Setting $W_1=K_{+}U_1$ and $W_2=U_2K_-$ with $U_1 U_2=U_r$, we
reproduce (\ref{san}) in the form (\ref{Sw1w2}). The corresponding
dreibein and Lorentz connection
\be \label{hKUK}
h=U_1^{-1}K_{+}^{-1}dK_{+}U_1-U_2K_-dK_-^{-1}U_2^{-1}+U_1^{-1}dU_1-U_2dU_2^{-1},
\ee
\be\label{oKUK}
\omega=U_1^{-1}K_{+}^{-1}dK_{+}U_1+U_2K_-dK_-^{-1}U_2^{-1}+U_1^{-1}dU_1+U_2dU_2^{-1}
\ee
do not depend on $t$, $\phi$ as soon as $U_{1,2}=U_{1,2}(r)$.
Therefore they remain well-defined in the BTZ case upon the identification
$\phi\to\phi+2\pi$.

It is convenient to use  the following matrices $U_{1,2}$
\begin{equation}
U_{1} = \Big(\f{A}{B}\Big)^{\f14}\left(
\begin{array}{cc}
0 & -\mu(r)\sqrt{B} \\
\eta(r)\sqrt{A} & \mu(r)\sqrt{A} \\
\end{array}
\right)\,,\quad U_{2} = \Big(\f{A}{B}\Big)^{\f14}\left(
\begin{array}{cc}
\mu(r) & 0 \\
-\mu^{-1}(r) & \eta(r)\sqrt{AB} \\
\end{array}
\right)\,,
\end{equation}
where $\mu(r)$, $\eta(r)$ are some functions that depend on
the radial coordinate and satisfy
\be\label{mu_eta}
\mu(r)\eta(r)=A^{-1}(r)\,.
\ee
The resulting matrices $W_{1\,\al}{}^{\gb}=(K_+U_1)_{\al}{}^{\gb}$,
$W_{2\,\al}{}^{\gb}=(U_2K_-)_{\al}{}^{\gb}$ are
\[
W_{1\,\al}{}^{\gb}=\sqrt{\f{u+x}{y-v}}\left(
\begin{array}{cc}
0 & -\mu (y-v) \\
\eta (u-x) & \mu (u-x)\\
\end{array}\right)\,,
\]
\be\label{W}
W_{2\,\al}{}^{\gb}=\sqrt{\f{u+x}{y-v}}\left(
\begin{array}{cc}
\mu & 0 \\
-\mu^{-1} & \eta (u-x)(y-v)  \\
\end{array}\right)\,.
\ee
According to \eqref{gcalfi2} and \eqref{gcalfi1}, the
corresponding dreibein and Lorentz connection have the form
\begin{align} \label{dreibein}
h_{11} &= A \mu^2 \left(-d\tilde{t} + d\tilde{\phi} + \f{1}{2AB}dA\right), \notag\\
h_{12} &= h_{21} =d\tilde{t} - \f{1}{2B}dA,\\
h_{22} &= -\mu^{-2} \left(d\tilde{t} + d\tilde{\phi} -
\f{1}{2AB}dA\right)\notag,
\end{align}
\begin{align} \label{lorcon}
\omega_{11} &= A \mu^2 \left(-d\tilde{t} + d\tilde{\phi} + \f{1}{2AB}dA\right), \notag\\
\omega_{12} &= \omega_{21} =-d\tilde{\phi} - \f{1}{2A}dA - \f{2}{\mu}d\mu,\\
\omega_{22} &= \mu^{-2} \left(d\tilde{t} + d\tilde{\phi} -
\f{1}{2AB}dA\right) \notag,
\end{align}
where $A, B$ and $\tilde{\phi}$, $\tilde{t}$ are defined in
\eqref{ABdef} and \eqref{tphitilde}, respectively. These expressions
are well-defined on $S^1$ with the cyclic coordinate
$\phi\sim\phi+2\pi$.

\section{Gauge function}\label{gaugefunc}

Locally, the equation \eqref{zero} admits a pure gauge solution
\be\label{gdg}
w_{0}(a,b|X)=-g^{-1}(a,b|X)\star dg(a,b|X)\,,
\ee
where $g(a,b|X)$ is some invertible ($g^{-1}\star g=g\star g^{-1}
=1$) element of the star-product algebra.
Once the gauge function $g(a,b|X)$ is known, in the unfolded
formulation this is equivalent to the full solution of the
linear problem. In particular, the global symmetry
parameters satisfying \eqref{eps} have the form
\be\label{epss}
\epsilon_{0}(a,b|X) = g^{-1}(a,b|X)\star\xi\star g(a,b|X)\,,
\ee
where $\xi=\xi(a,b)$ is an arbitrary $X$-independent element of the
star-product algebra. In Section \ref{Fockmod} it is explained how
the knowledge of $g(a,b|X)$ allows one to reconstruct a generic
solution of free field equations. As is well-known, the pure gauge
representation \eqref{gdg} is invariant under left global
transformations
\be
\label{gleft}
g(a,b|X)\to f(a,b)\star g(a,b|X)
\ee
with an $X$-independent star-invertible $f(a,b)$.

Using the results of \cite{DV} we obtain that the
gauge function $g(a,b|W_1,W_2)$, that generates \eqref{gcalfi2}
and \eqref{gcalfi1} via \eqref{gdg}, is
\begin{align}\label{gWfunc}
&g(a,b|W_1,W_2)=\f{4}{\sqrt{\det||(W_1+1)(W_2+1)||}}\exp\Big(
-\f12\Pi^{\al\gb}(W_1)T^{+}_{\al\gb}-
\f12\Pi^{\al\gb}(W_2)T^{-}_{\al\gb}\Big)
, \notag \\
&g^{-1}(a,b|W_1,W_2)=\f{4}{\sqrt{\det||(W_1+1)(W_2+1)||}}\exp\Big(
\f12\Pi^{\al\gb}(W_1)T^{+}_{\al\gb}+
\f12\Pi^{\al\gb}(W_2)T^{-}_{\al\gb}\Big)
\end{align}
with\footnote{A matrix fraction $\f{B}{A}$ is understood as
$A^{-1}B$. Note that \eqref{pi} is an analogue of the so-called
Caley's transformation.} \be\label{pi}
\Pi_{\al\gb}(W)=\Pi_{\gb\al}(W)=\Big(\f{W-1}{W+1}\Big)_{\al\gb}
\ee and \be T^{\pm}_{\al\gb}=a_\al a_\gb + b_\al b_\gb \pm (a_\al
b_\gb + b_\al a_\gb). \ee Here $T^\pm_{\alpha\beta}$ are the
generators of the $sp(2)$ subalgebras of $sp(2)\oplus sp(2)$
generated by the two mutually commuting sets of oscillators
$\alpha^{\pm}_\alpha = a_\alpha \pm b_\alpha$ satisfying the
commutation relations $[\alpha^{\pm}_\alpha,
\alpha^{\pm}_\beta]_\star = \pm 2\epsilon_{\alpha\beta}$.  In
practice, it is often convenient to use the star-product defined
in terms of mutually  commuting oscillators $\ga_\beta^{\pm}$
as\footnote{Note that linear transformations of the generating
elements of the Weyl star-product form automorphisms of the
star-product algebra. This is the consequence of the definition of
the Weyl star-product as resulting from the totally symmetrized
ordering prescription in terms of the generating oscillators,
which is insensitive to the particular choice of basis
oscillators.} \be \label{newsp} (f*g)(\al^{\pm})=\f
{1}{(2\pi)^2}\int f(\al^{\pm}+u)g(\al^{\pm}+v)e^{\mp
u_{\al}v^{\al}}d^2ud^2v\,. \ee Taking into account that
$T^\pm_{\alpha\beta}=\alpha^\pm_\alpha\alpha^\pm_\beta$, the
following useful formula for the gauge function \eqref{gWfunc}
results from \eqref{newsp} (for more detail see \cite{DV})
\be\label{gom} g(a,b|K_1, K_2)\star
g(a,b|U_1,U_2)=g(a,b|K_1U_1,U_2K_2)\, \ee at the condition that
the matrices $K_{1,2}+1$ and $U_{1,2}+1$ are non-degenerate. Owing
to the equality
\[
\Pi_{\al\gb}(W)=-\Pi_{\al\gb}(W^{-1})\,,
\]
 the transformation of the gauge function \eqref{gWfunc}
\be\label{lorg} g(a,b|W_1, W_2)\to g(a,b|W_1, W_2)\star
\Lambda^{-1}(a,b|V) \ee with
\begin{align}\label{lorlambda}
\Lambda(a,b|V)&=\f{4}{\det||V+1||}\exp\Big(\f12\Pi^{\al\gb}(V)(T^{+}_{\al\gb}-T^{-}_{\al\gb})\Big), \notag \\
\Lambda^{-1}(a,b|V)&=\f{4}{\det||V+1||}\exp\Big(-\f12\Pi^{\al\gb}(V)(T^{+}_{\al\gb}-T^{-}_{\al\gb})\Big),
\end{align}
where $V_{\al\gb}(X) \in Sp(2)$, describes the local Lorentz
transformation of the dreibein \eqref{gcalfi2}
\be\label{lortrans}
h_{\al\gb}\to V^{\gga}{}_{\al}V^{\gd}{}_{\gb}h_{\gga\gd}\,,
\ee
that leaves invariant the metric \eqref{hh}.

Also from \eqref{gom} it follows that the dreibein \eqref{hKUK} and
Lorentz connection \eqref{oKUK} are reproduced by the gauge function
of the form
\be\label{decomp}
g(a,b|t,\phi,r)=\Phi(a,b|t,\phi)\star U(a,b|r)
\ee
with
$$
\Phi(a,b|t,\phi)=\f{4}{\sqrt{\det||(K_++1)(K_-+1)||}}\exp\Big(
-\f12\Pi^{\al\gb}(K_{+})T^{+}_{\al\gb}-
\f12\Pi^{\al\gb}(K_-)T^{-}_{\al\gb}\Big)\,,
$$
$$
U(a,b|r)=\f{4}{\sqrt{\det||(U_1+1)(U_2+1)||}}\exp\Big(
-\f12\Pi^{\al\gb}(U_1)T^{+}_{\al\gb}-
\f12\Pi^{\al\gb}(U_2)T^{-}_{\al\gb}\Big)\,,
$$
provided that $U_{1}U_{2}=U_r$ \eqref{KUK}.

Note that the metric (\ref{hh}) is invariant under global left and
right group multiplications of $S_{\gamma}{}^{\delta}(X)$ $ S\to H
S \tilde{H}, $ where $H$ and $\tilde{H}$ are some $X$-independent
elements of $Sp(2)$. We will use this ambiguity in Section
\ref{Fockmod} to analyze the problem away from the outer horizon.
For that purpose let us choose
\be\label{global}
S_{\gga\gd}=(HS_0)_{\gga\gd}
\ee
 with
the constant matrix $H$ of the form
\begin{equation}
    H_{\gga}{}^{\gd} = \left(
      \begin{array}{cc}
        \al & \gb \\
        \gb & \al \\
      \end{array}
    \right)\,,
\end{equation}
where $\al^2=A(r_0)$ and $\gb^2=B(r_0)$ for some $r_0>r_{+}$. From
\eqref{ABdef} it follows that $\al^2-\gb^2=1$. The new matrix
$S_{\gamma\delta}$ is
\begin{equation} \label{newSm}
    S_{\gga\gd} = \left(
      \begin{array}{cc}
        \al (y-v)+\gb(x-u) & \al(x+u)-\gb(y+v) \\
        \gb (y-v)+\al(x-u) & \gb(x+u)-\al(y+v) \\
      \end{array}
    \right).
\end{equation}
Taking into account \eqref{Sw1w2} and \eqref{gom}, this transformation
is reached by
\be\label{redef}
g(a,b|W_1,W_2)\to K(a,b|H)\star g(a,b|W_1,W_2)\,,
\ee
where
\[
K(a,b|H)=\f{2}{\sqrt{\det||H+1||}}e^{
-\f12\Pi^{\gamma\delta}(H)T^{+}_{\gamma\delta}}\,.
\]
Thus, dreibein and Lorentz connection remain invariant under the
transformation \eqref{global}.

Note, that $K(a,b|1)=1$. As  explained in the next section, the case
with $K(a,b|H)\neq 1$ plays a role of  regularization that allows us
to analyze the problem away from a point where a solution of
interest develops singularity. After the solution is found we remove
the regularization by setting  $\al=1$, $\gb=0$.

\section{Unfolded  equations for $3d$ massless fields and Fock
module}\label{Fockmod}

To formulate free dynamical equations for massless fields in the BTZ
black hole background we follow the unfolded formulation of the
massless field equations worked out in \cite{V91,SV,DV}.
In particular, as shown
in \cite{SV}, the free field dynamics of massless spins $s=0$ and
$s=\f12$ in $AdS_3$ can be formulated  in a manifestly conformal
invariant way in terms of sections of a certain Fock fiber bundle.
Namely, consider space-time fields that take values in the Fock
module generated by the oscillator $b^{\al}$
\be \label{fockdef}
|C(b|X)\rangle = C(b|X) \star |0\rangle\langle 0|,
\ee
where $C(b|X)$ is the generating function
\be \label{multiplet}
C(b|X)=\sum_{k=0}^\infty \frac{1}{k!}C_{\alpha_1\ldots
\alpha_k}(X)b^{\alpha_1}\ldots b^{\alpha_k},
\ee
and $|0\rangle\langle 0|=e^{-2a_\gamma b^\gamma}$ is the Fock vacuum
satisfying
\be
a_{\al}\star\vac=0\,,\qquad \vac\star b_{\al}=0\,.
\ee
The dynamical massless scalar and spinor fields  identify with the
lowest  components
\be \label{dynfields}
C(X)=\left.C(b|X)\right|_{b=0}\,,\qquad C_{\al}(X)=\f{\p}{\p
b^{\al}}C(b|X)\Big|_{b=0}\,.
\ee
The dynamical equations for massless fields in a locally $AdS_3$
space-time can be formulated in the unfolded form
\begin{equation} \label{Ceq}
d|C(b|X)\rangle - w_0(a,b|X) \star |C(b|X)\rangle = 0\,,
\end{equation}
where $w_{0}(a,b|X)$ fulfils the zero-curvature condition
\eqref{zero}. Let us show that (\ref{Ceq}) is equivalent  to
conformal Klein-Gordon and Dirac equations along with constraints
that express higher multispinor components in the expansion
(\ref{multiplet}) via higher derivatives of the dynamical fields
\cite{SV}. Using  (\ref{connect}), the equation (\ref{Ceq}) can be
rewritten in the component form as
\begin{equation}\label{Ceqtens}
   DC_{\al_1\dots \al_k}=\frac{k(k-1)}{4}h_{(\al_1 \al_2}C_{\al_3\ldots \al_k)}+
   \frac14 h^{\gb\gl}C_{\gb\gl \al_1\ldots \al_k}\,,
\end{equation}
where parentheses denote total symmetrization and $D$ is the Lorentz
covariant differential
\[
DC_{\al_1\ldots \al_k}=dC_{\al_1\ldots
\al_k}+\frac{k}{2}\omega_{(\al_1}{}^\gga C_{\gga \al_2\ldots
\al_k)}\,.
\]
Setting $k=0$ and $k=2$ one gets from (\ref{Ceqtens})
\begin{eqnarray} \label{scalchain1}
 D_n C &=& \frac14 h_{n,}{}^{\alpha\beta}C_{\alpha\beta}, \\
 D_n C_{\alpha\beta} &=& \frac12 h_{n,\alpha\beta}C +\frac14
 h_{n,}{}^{\gamma\delta}C_{\alpha\beta\gamma\delta}\,.
 \label{scalchain2}
\end{eqnarray}
Using that $C_{\alpha\beta\gamma\delta}$ is symmetric in its indices
we obtain from  \eqref{scalchain1}, \eqref{scalchain2} the  Klein-Gordon
equation for the scalar field $C(X)$
\begin{equation}\label{KlGn}
   \Box C \equiv D^nD_n C=\frac34 C\,.
\end{equation}
Analogously, from the equations (\ref{Ceqtens}) with $k=1$
 we obtain the Dirac equation
\be\label{dirac}
h^{n}{}_{,\,\al\gb}D_{n}C^{\gb}=0\,.
\ee
All other fields in the multiplet \eqref{multiplet} are
expressed by (\ref{Ceqtens}) via derivatives of the dynamical fields \eqref{dynfields}.

The gauge transformation \eqref{gauge} acts on the Fock module in a
natural way
\be\label{gaugeFock}
\gd\Fock=\gep(a,b|X)\star\Fock\,.
\ee
In particular, the Lorentz transformation \eqref{lorg}
of the gauge function acts as follows
\be\label{lorFock}
\Fock\to \Lambda(a,b|V)\star
|C(b_{\ga}|X)\rangle=|C(V_{\ga}{}^{\gb}b_{\gb}|X)\rangle\,,
\ee
where $\Lambda(a,b|V)$ is defined in \eqref{lorlambda}.

Choosing $w_{0}(a,b|X)$ in the pure gauge form
\eqref{gdg}, one obtains general local solution for $\Fock$
in the form
\be\label{Fock}
\Fock=g^{-1}(a,b|X)\star|C(b|X_{0})\rangle=g^{-1}(a,b|X)\star
C(b)\star\vac\,,
\ee
where $|C(b|X_0)\rangle =C(b)\star\vac $ plays a role of initial
data. The meaning of the formula \eqref{Fock} is that, for
$g(a,b|X_0)=1$ at some $X=X_0$,  it gives a covariantized Taylor
expansion that reconstructs a solution in terms of its on-shell
nontrivial derivatives at $X=X_0$ parameterized by $C(b)$. Note
that this interpretation can be adjusted to any given regular point $X_0$
by the redefinition of the gauge function
\be
\label{red}
g(a,b|X) \to \tilde{g}(a,b|X) = g^{-1}(a,b|X_0)\star g(a,b|X)
\ee
that leaves unchanged the flat connection \eqref{gcalfi2} and
\eqref{gcalfi1}, effectively implying the redefinition of the $C(b)$
$$
\Fock= \tilde{g}^{-1}(a,b|X) \star \tilde{C}(b)\star\vac\,,\qquad
\tilde{C}(b)\star\vac = g^{-1}(a,b|X_0)\star C(b)\star\vac \,.
$$
Clearly, this formalism cannot be applied to a point $X_0$ at
which a solution $C(b|X)$ develops a singularity. In practice, a
space-time singularity at $X_0$ manifests itself in the
nonexistence of the corresponding $\tilde{C}(b)\star\vac$ (note
that the star-product of nonpolynomial functions is not
necessarily well defined). The way out is to perform some
redefinition (\ref{red}) that would correspond to the analysis at
some regular point of the solution.

The unfolded form of massless field equations (\ref{Ceqtens}) is
manifestly conformal invariant with the $3d$ conformal algebra
$sp(4)\sim o(3,2)$ generated by various bilinears of the
oscillators (\ref{osci}). It can be extended to the massive case by the
replacement of the usual oscillators $a_\alpha, b^\alpha$ with the
so-called deformed oscillators along the lines of \cite{PV} (and
references therein) or using the Fock module realization of the
deformed oscillator algebra with the doubled number of oscillators
as in \cite{SV} and in this paper. (Note that, as expected, the
conformal algebra $sp(4)$ breaks down to the $AdS_3$ algebra
$sp(2)\oplus sp(2)$ in the massive case because of the properties
of the deformed oscillators.) As the corresponding formulation is
technically more involved the case of arbitrary mass is not
considered in this paper.

In the standard formulation, the case of a massive scalar field in
the BTZ black hole background \eqref{BTZ} was originally considered
in \cite{scalarBTZ1,scalarBTZ2}. A solution of
\[
\Box C=m^2C
\]
with definite energy  $E$  and angular momentum $L$ has the form
\be\label{scalar}
C(t,r,\phi)=e^{-iEt}e^{iL\phi}R(r)\,,
\ee
where
\be \label{Rscalar}
R(r)=\left(1-A(r)^{-1}\right)^{\frac{P+Q}{2}}A(r)^{-\gamma}f(r)\,,
\ee
\be \label{PQ}
P=i\frac{E-L}{2(r_+-r_-)}\,,\qquad
Q=i\frac{E+L}{2(r_++r_-)}\,,\qquad m^2=4\gamma(1-\gamma)
\ee
and
\[
f(r)=K_1 F(P+\gamma,Q+\gamma,2\gamma;A(r)^{-1}) + K_2 A(r)^{2\gamma-1}
F(P+1-\gamma,Q+1-\gamma,2-2\gamma;A(r)^{-1})\,
\]
with  $K_1$, $K_2$ being integration constants. $A(r)$ is defined
in \eqref{ABdef} and $F(a,b,c;x)$ is the hypergeometric function.
Note that the substitution $\gamma\to 1-\gamma$ interchanges the
two independent basis solutions. The massless case \eqref{KlGn}
corresponds to $m^2=3/4$ and, consequently, $\gamma=1/4$.

In the rest of this paper we show how known results for massless
scalar and spinor fields in BTZ black hole background
are reproduced in our approach. To single
out the states with definite energy and angular momentum in the
multiplet $\Fock$ we impose the following conditions
\be\label{eigenval}
\gep_t\star\Fock=-i E\Fock\,,\quad \gep_{\phi}\star\Fock=i L\Fock\,,
\ee
where $\gep_t=g^{-1}\star\xi_t\star g$ ,
$\gep_{\phi}=g^{-1}\star\xi_{\phi}\star g$  are the symmetry generators
of the BTZ Killing vectors $\f{\p}{\p\phi}$ \eqref{phikil} and
$\f{\p}{\p t}$ \eqref{tkil}. Using \eqref{Fock} we rewrite
\eqref{eigenval} as
\be\label{eigenval2}
\xi_t\star C(b)\star\vac=-i E C(b)\star\vac\,,\qquad
\xi_{\phi}\star C(b)\star\vac=i L C(b)\star\vac\,.
\ee
To analyse these equations, which define initial data $C(b)$, we
have to find the form of the generators $\xi_t$ and $\xi_\phi$ in
the star-product algebra. This is done in the next section.

The following comment is now in order. The solution \eqref{scalar}
is singular at $r=r_{+}$. Therefore, it cannot be treated in the
unfolded formulation within the expansion at the horizon.
Indeed, in Section
\ref{solutions} we will see that the equations \eqref{eigenval2},
that correspond to the expansion at $r=r_+$, admit no solutions
with $C(b)$ regular in $b^\alpha$, that can be interpreted in terms of
the Fock module. Note,
that since the gauge function \eqref{gWfunc} is regular on the
horizon, this means that the singularity of a solution results
from  the condition that it carries definite energy and momentum
and can be avoided by relaxing this condition.

To see that the gauge function \eqref{gWfunc} indeed corresponds to the
expansion near horizon we observe that it can be transformed to unity by
a Lorentz transformation. Actually, according to
\eqref{lorg} and \eqref{Fock} the Lorentz transformation
$\Lambda(a,b|W_2)$ acts on $g(a,b|W_1,W_2)$ as
\[
\tilde{g}(a,b|W_1, W_2)=g(a,b|W_1, W_2)\star
\Lambda^{-1}(a,b|W_2)=g(a,b|W_1 W_2,1)
\]
and thus,
\[
\tilde{g}(a,b|W_1(X_0), W_2(X_0))=1\quad \textnormal{iff}\quad
W_1(X_0) W_{2}(X_0)=1\,.
\]
The choice of the gauge function \eqref{redef} with
$H_{\gamma}{}^{\delta}=\gd_{\gamma}{}^{\delta}$ corresponds to
$S_{\gamma}{}^{\delta}=(HW_1W_2)_{\gamma}{}^{\delta}=\gd_{\gamma}{}^{\delta}$
at the point $X_{0}=\{r=r_{+}, t=0, \phi=0\}$ that belongs to the
horizon. Indeed,
$S_0(X_0)_{\gamma}{}^{\delta}=\gd_{\gamma}{}^{\delta}$ implies
$v_0=x_0=y_0=0$, $u_0=1$ that corresponds to  $r_0=r_{+}$,
$t_0=\phi_0=0$.
To avoid this problem we apply the transformation \eqref{redef} to
achieve  the redefinition \eqref{newSm}. Now
$S(X_0)_{\gamma}{}^{\delta}=\gd_{\gamma}{}^{\delta}$ at the point
$X_0=\{r_0>r_+, t=0, \phi=0\}$ which, unless $\al=1, \gb=0$, is
regular thus allowing consistent unfolded analysis at least in
some its neighbourhood.
The regularization with $\al\neq 1$ and $\gb \neq 0$ is necessary
for intermediate calculations (see Appendix B) while
the limit $\al\to 1$, $\gb\to 0$ can be taken  in the final
expression for $C(b|X)$.
Recall, that the ambiguity in $H_{\gamma}{}^{\delta}$ does not
affect  BTZ black hole connections \eqref{dreibein}, \eqref{lorcon}.

\section{Star-product realization of $AdS_3$ Killing
vectors}\label{killings}

Any Killing vector $\f{\p}{\p\zeta}$  of $AdS_3$ is a linear
combination of $J_{ab}$  \eqref{momentum}, i.e.,
\be
\f{\p}{\p\zeta}=\Omega^{ab}J_{ab}\,,
\ee
where $\Omega^{ab}=-\Omega^{ba}$ are some constants. In the
star-product algebra it corresponds to a global symmetry generator
$\xi$ that belongs to $sp(2)\oplus sp(2)$ algebra, i.e.,
\be
\xi=(\kappa_{1})^{\al\gb}L_{\al\gb}+(\kappa_{2})^{\al\gb}P_{\al\gb}
\ee
with some constant matrices $\kappa_1$ and $\kappa_2$. To find
$\xi_t$ and $\xi_{\phi}$ associated with the  BTZ Killings
$\f{\p}{\p t}$ and $\f{\p}{\p\phi}$ let us evaluate the on-shell
action of the  generators $L_{\al\gb}$ and $P_{\al\gb}$ on the
scalar field. We will use the gauge function \eqref{redef} with
$S_{\al\gb}$  \eqref{newSm}.

Let us introduce the generating parameters $
\xi^{L}=(\kappa_1)^{\ga\gb}L_{\ga\gb}$ for a Lorentz generator and
$\xi^{P}=(\kappa_2)^{\ga\gb}P_{\ga\gb}$ for a $AdS$-translation
generator. Using \eqref{epss}, \eqref{gWfunc}, \eqref{gaugeFock} and
the equations of motion  it is not hard to obtain (see Appendix A
for details)
\be\label{vecsym}
\gd^{L}C(X)=\frac12(\kappa_1)^{\ga\gb}\mathcal{L}_{\alpha\beta,}{}^{n}
\p_n C(X)\,
\ee
and
\be\label{vecsym2}
\gd^{P}C(X)=(\kappa_2)^{\ga\gb}\mathcal{P}_{\alpha\beta,}{}^{n}\p_n
C(X)\,,
\ee
where
\begin{equation}\label{PLfields}
\mathcal{P}_{\alpha\beta,n}= \p_n S_{\alpha\gamma}
S_\beta{}^\gamma - \p_n S_{\gamma\alpha} S^\gamma{}_\beta\,, \quad
\mathcal{L}_{\alpha\beta,n}=\frac12(\p_n S_{\alpha\gamma}
S_\beta{}^\gamma + \p_n S_{\gamma\alpha} S^\gamma{}_\beta)\,.
\end{equation}
Substituting \eqref{newSm} into (\ref{PLfields}) and comparing the
resulting expression with the $AdS$ Killing vectors \eqref{momentum}
we obtain
$$
\mathcal{L}_{\gamma\delta}= \left(
 \begin{array}{cc}
   \alpha\beta(J_{12}-J_{03})-\alpha^2J_{23}-\beta^2J_{01}+J_{02}
   & -\alpha\beta(J_{01}+J_{23})-\alpha^2J_{03}+\beta^2J_{12} \\
   -\alpha\beta(J_{01}+J_{23})-\alpha^2J_{03}+\beta^2J_{12}
   & \alpha\beta(J_{12}-J_{03})-\alpha^2J_{23}-\beta^2J_{01}-J_{02} \\
 \end{array}
\right),
$$
\be\label{starkill}
\mathcal{P}_{\gamma\delta}=2 \left(
 \begin{array}{cc}
   \alpha\beta(J_{12}-J_{03})-\beta^2J_{23}-\alpha^2J_{01}+J_{13}
   & -\alpha\beta(J_{01}+J_{23})-\beta^2J_{03}+\alpha^2J_{12} \\
   -\alpha\beta(J_{01}+J_{23})-\beta^2J_{03}+\alpha^2J_{12}
   & \alpha\beta(J_{12}-J_{03})-\beta^2J_{23}-\alpha^2J_{01}-J_{13} \\
 \end{array}
\right).
\ee
From here we find that the components $J_{03}$ and $J_{12}$ which
contribute to the  BTZ Killing vectors \eqref{phikil} and
\eqref{tkil} are
\begin{eqnarray} \label{J12}
 J_{03}&=& -\frac14\tau_1^{\gamma\delta}\mathcal{P}_{\gamma\delta}
 - \frac12\tau_2^{\gamma\delta}\mathcal{L}_{\gamma\delta}\,, \\
 J_{12}&=&\frac14\tau_2^{\gamma\delta}\mathcal{P}_{\gamma\delta}
 + \frac12\tau_1^{\gamma\delta}\mathcal{L}_{\gamma\delta}\,,
\end{eqnarray}
where
\be \label{taumat}
\tau_1^{\gga\gd}=\left(
 \begin{array}{cc}
   -\alpha\beta & \beta^2 \\
   \beta^2 & -\alpha\beta \\
 \end{array}
\right), \qquad \tau_2^{\gga\gd}=\left(
 \begin{array}{cc}
   -\alpha\beta & \alpha^2 \\
   \alpha^2 & -\alpha\beta \\
 \end{array}
\right).
\ee
Note that the matrices $\tau_1$ and $\tau_2$ satisfy
\[
\f12\tau_{1\,\gga\gd}\tau_1^{\gga\gd}=\gb^2,\quad
\f12\tau_{2\,\gga\gd}\tau_{2}^{\gga\gd}=\al^2,\quad
\tau_{1\,\gga\gd}\tau_{2}^{\gga\gd}=0,
\]
\[
\tau_{2}^{\gga\gd}-\tau_{1}^{\gga\gd}=\left(
\begin{array}{cc}
0 & 1 \\
1 & 0 \\
\end{array}\right).
\]
Thus, the oscillator realization of BTZ Killing vectors on the
Fock module is
\begin{eqnarray}
\xi_t &=& \frac12(r_-\tau_1^{\gga\gd}
+r_+\tau_2^{\gga\gd})L_{\gga\gd} +\frac14(r_+\tau_1^{\gga\gd}
+r_-\tau_2^{\gga\gd})P_{\gga\gd}\,,\label{osckilt} \\
\xi_\phi &=& -\frac12(r_+\tau_1^{\gga\gd}
+r_-\tau_2^{\gga\gd})L_{\gga\gd} -\frac14(r_-\tau_1^{\gga\gd}
+r_+\tau_2^{\gga\gd})P_{\gga\gd}\,. \label{osckilf}
\end{eqnarray}

\section{Explicit solutions for massless fields}\label{solutions}

Having found the oscillator realization of BTZ Killing vectors
\eqref{osckilt}, \eqref{osckilf}, we can rewrite the equations
\eqref{eigenval2} on the generating function for a field with
definite energy and angular momentum in the following form
\begin{subequations} \label{mainsys}
\begin{eqnarray} \label{eigensum}
(\tau_2-\tau_1)^{\gga\gd}(L-\frac12 P)_{\gga\gd}
\star C(b)\star\vac&=&-4P C(b)\star\vac\,, \\
(\tau_2+\tau_1)^{\gga\gd}(L+\frac12 P)_{\gga\gd} \star
C(b)\star\vac&=&-4Q C(b)\star\vac\,,
\end{eqnarray}
\end{subequations}
where $P$ and $Q$ are given in \eqref{PQ}.
Let
\be
b^{\al}=(p,q)\,.
\ee
Then the  system (\ref{mainsys})
amounts  to the two
second-order differential equations
\be\label{firsteq}
(p\,\partial_p-q\,\partial_q+
p\,q-\partial_p\partial_q)C(p,q)=-4PC(p,q),
\ee
\begin{align} \label{secondeq}
&-\alpha\beta\left(\partial_p\partial_p+\partial_q\partial_q+p^2
+q^2+2p\,\partial_q-2q\,\partial_p\right)C(p,q) \notag \\
&+(\alpha^2+\beta^2)\left(p\,\partial_p-q\,\partial_q+\partial_p\partial_q-
p\,q\right)C(p,q) =-4QC(p,q).
\end{align}
Note that  the  case of $\al=1$ and $\gb=0$
is degenerate reducing the sum of the equations \eqref{firsteq}
and (\ref{secondeq})  to the first order equation. As a
result, the system  (\ref{firsteq}), (\ref{secondeq}) at
$\al=1$, $\gb=0$ admits no solutions regular
in $b^\alpha$. Indeed, in this case from \eqref{firsteq}
and \eqref{secondeq} it follows that
\[
(p\p_p-q\p_q)C(p,q)=-2(P+Q)C(p,q)
\]
and, therefore, $C(p,q)=p^{-2(P+Q)}\chi(pq)$ is not regular in the
oscillators $b^{\al}$ for physical values of
$P$ and $Q$.

The substitution  $C(p,q)=e^{p\,q}f(p,q)$ reduces (\ref{firsteq})
to
\begin{equation}
(\partial_p\partial_q+2q\,\partial_q)f(p,q)=(4P-1)f(p,q),
\end{equation}
which can be solved as
\begin{equation}\label{intsol}
f(p,q)=\int\limits_{-\infty}^{\infty}
e^{-\frac{\alpha}{4\beta}s^2}g(s) e^{p
s}(\frac{s}{2}+q)^{2P-\frac12}ds\,,
\end{equation}
where $g(s)$ is still arbitrary. Plugging this  into
\eqref{secondeq} leads to the differential equation for
$g(s)$
\be\label{hhf}
\al\gb g''(s)-\f12 s g'(s)-(Q+\f14)g(s)=0,
\ee
which is the confluent hypergeometric equation. Its general
solution can be expressed in the integral form as a superposition
of the following two basis solutions
\be
\int\limits_{0}^{\infty}w^{2Q-\f12}e^{-\al\gb w^2+sw}d w\qquad
\textnormal{and}\qquad\int\limits^{\infty}_{0}w^{2Q-\f12}e^{-\al\gb
w^2-sw}d w\,.
\ee
The integrals are convergent since $\al\gb>0$ and $Re\,
Q>-\f14$.

Abusing notation, we denote general solution of \eqref{hhf} as
\be\label{gensol}
\int w^{2Q-\f12}e^{-\al\gb w^2+sw}d w\,,
\ee
assuming by this a linear combination of the integrals
\be\label{inthhf}
\int\limits_{0}^{\infty}w^{2Q-\f12}e^{-\al\gb w^2+sw}d w\qquad
\textnormal{and}\qquad\int\limits_{-\infty}^{0}w^{2Q-\f12}e^{-\al\gb
w^2+sw}d w\,.
\ee
Note that although the second integral in \eqref{inthhf} is
infinitely-valued, the ambiguity  is modulo an arbitrary constant
phase factor that can be absorbed into an integration constant.

Using $g(s)$ from \eqref{gensol} and changing the integration
variable $s\to s-2q$ in \eqref{intsol}, we obtain the generating
function in the form
\begin{align} \label{Cb}
C(p,q)&=e^{-p\,q}\int\limits_{-\infty}^{\infty}ds\int dw\,
e^{-\f{\al}{4\gb}(s-2q)^2-\al\gb
w^2+(s-2q)w+sp}s^{2P-\f12}w^{2Q-\f12} \notag \\
&=\int\limits_{-\infty}^{\infty}ds\int dw\,
e^{m_{\gga\gd}b^{\gga}b^{\gd}+n_{\gga}b^{\gga}} e^{-\alpha\beta
w^2+sw-\frac{\ga}{4\gb}s^2} s^{2P-\f12}w^{2Q-\f12} \,,
\end{align}
where we use the notation
\[
m_{\gga\gd} = \left(
                  \begin{array}{cc}
                    0 & -\frac12 \\
                    -\frac12 & -\frac{\alpha}{\beta} \\
                  \end{array}
                \right),\qquad n_{\gga}=(s,
                \frac{\al}{\gb}s-2w).
\]
Using \eqref{Cb} and \eqref{redef} one can calculate generating
function \eqref{Fock}, redefining the integration variable $\gb
w\to w$ and setting then $\al=1$, $\gb=0$. We obtain the following
integral representation for the generating function $C(b|X)$ up to
a constant factor (see Appendix B)

\begin{align} \label{CbBTZ}
C(b|t,r,\phi)&=e^{-iEt}e^{iL\phi}
A(r)^{-Q-\frac12}(1-A(r)^{-1})^{\frac{P+Q}{2}} e^{- b^1
b^2}\int\limits_{-\infty}^{\infty}ds\int dw \,
s^{2P-\f12}w^{2Q-\f12}\notag  \\
& \times \exp\left(-\f{s^2}{4}-\f{w^2}{4A(r)}+\f{sw}{2A(r)}
+\mu(r) s b^1 - \eta(r) w b^2\right),
\end{align}
where $A(r)$ and $\mu(r),\eta(r)$ are defined in \eqref{ABdef} and
\eqref{mu_eta}, respectively. Note that, as discussed in Section
\ref{Fockmod} and in the beginning of this section, the formalism
does not allow to set $\al=1, \gb=0$ in $C(b)$ before completing
its star multiplication with $g^{-1}(a,b|W_1,W_2)$.

By construction, the generating function (\ref{CbBTZ})
gives solutions of free massless equations
in the BTZ background along with all derivatives of the massless
fields as coefficients of the expansion in powers of $b^\ga$.
Using the standard integral representation for the
hypergeometric function (see, e.g., \cite{ryzhik})
the generating function (\ref{CbBTZ}) can be written in the form
\begin{align} \label{genryad}
&C(b|t,r,\phi) =e^{-iEt}e^{iL\phi}
(1-A^{-1})^{\frac{P+Q}{2}}A^{-\frac14}
\sum_{m=0}^{\infty}\sum_{n=0}^{\infty} \f{\mu^m (-\eta)^n}{m!n!} A^{\f{n}{2}}
(b^1)^m (b^2)^n e^{-b^1b^2}\notag \\
&\times \left[ K_1
F\left(P+\f{2m+1}{4},Q+\f{2n+1}{4},\frac12;A^{-1}\right) + K_2
A^{-\frac12}F\left(P+\f{2m+3}{4},Q+\f{2n+3}{4},\frac32;A^{-1}\right)
\right],
\end{align}
where $K_1, K_2$ are arbitrary integration constants and $P,Q$ are
defined in \eqref{PQ}.

As explained in Section \ref{Fockmod}, a solution for the scalar
field is given by $C(0|X)$ \eqref{dynfields}. So, from
\eqref{genryad} we obtain
\begin{align} \label{solscal}
C(t,r,\phi) &=e^{-iEt}e^{iL\phi}\left(1-A^{-1}\right)^{\frac{P+Q}{2}}A^{-
\frac14} \notag \\
&\times \left[ K_1
F\left(P+\frac14,Q+\frac14,\frac12;A^{-1}\right) + K_2
A^{-\frac12}F\left(P+\frac34,Q+\frac34,\frac32;A^{-1}\right)
\right],
\end{align}
This formula coincides with the solution of massless scalar equation \eqref{KlGn} in the
BTZ background originally obtained in \cite{scalarBTZ1,scalarBTZ2}
for any value of mass.

Also it is now straightforward to find from \eqref{genryad} the
solution for the spinor field $C_\al (X)$ \eqref{dynfields}
\be \label{solspin}
C_{\al}(t,r,\phi)=
e^{-iEt}e^{iL\phi}\left(1-A^{-1}\right)^{\frac{P+Q}{2}}A^{-\frac14}
(K_1 \,\psi_{1\alpha} + K_2 \,\psi_{2\alpha} ),
\ee
with
\[
\psi_{1}=\left(
\begin{array}{cc}
\mu F\left(P+\f{3}{4},Q+\f{1}{4},\f{1}{2};A^{-1}\right) \\
-(Q+\f14)\eta F\left(P+\f{3}{4},Q+\f{5}{4},\f{3}{2};A^{-1}\right)\\
\end{array}\right)
\]
and
\[
\psi_{2}=\left(
\begin{array}{cc}
(P+\f14)\mu A^{-\f12}F\left(P+\f54,Q+\f34,\f{3}{2};A^{-1}\right) \\
-\eta A^{\f12}F\left(P+\f14,Q+\f34,\f12;A^{-1}\right)\\
\end{array}\right).
\]
Different  choices of the functions $\mu(r), \eta(r)$ \eqref{mu_eta}
correspond to different Lorentz gauges in the general solution
of Dirac equation \eqref{dirac} with definite energy $E$ and
angular momentum $L$ in BTZ black hole background. Note that our
Lorentz gauge differs from that of \cite{BTZspinor,BSS}.

\section{Extremal BTZ black hole}\label{extreme}

Exact solutions for the Klein-Gordon and Dirac equations in the
extremal BTZ background were found in \cite{extr1} and
\cite{extr2}. In the extremal case with $M=|J|$ the two horizons
coincide and the parametrization \eqref{btzcoord} cannot be used.
As before, black hole connection  $w_0(a,b|X)$ is expressed via
the gauge function $g(a,b|W_1,W_2)$ but now the ambient
coordinates $X^a$ are parameterized differently (see \cite{BHTZ}).
In the extremal case the Killing vector responsible for the
identification \eqref{fact} has additional terms that cannot be
removed by a $SO(2,2)$ transformation
\be\label{extrkill}
\f{\p}{\p \phi} = -\gl r_+ J_{12} + \gl r_- J_{03} + J_{13} -
J_{23}.
\ee
Consequently, the system of equations \eqref{mainsys} changes its
form. Fortunately, to obtain the solutions in the extremal case it
is not necessary to solve the equations again. As pointed out e.g. in
\cite{troost} one can simply take
the limit for the solutions \eqref{solscal} and \eqref{solspin}. Namely,
$$
A^{-1}P = \kappa = i\f{(E-L)(r_++r_-)}{2(r^2-r^2_-)}
$$
is regular in the limit $r_+ \rightarrow r_-$ ($P$ is defined in
\eqref{PQ}). Now let us substitute $ A^{-1}=\f{\kappa}{P}$ in
\eqref{solscal}, \eqref{solspin} and consider the limit $r_+
\rightarrow r_-$ or, equivalently, $P \rightarrow \infty$. The
result can be written in terms of Whittaker functions $M_{p,q}(x)$
\cite{Vilenkin}.

For the massless scalar we obtain
\be
C(t,r,\phi) = e^{-iEt}e^{iL\phi} \left(K_1 M_{-Q,-\f14}(\kappa_e) +
K_2 M_{-Q,\f14}(\kappa_e)\right),
\ee
where
\be
 \kappa_e=i\f{(E-L)\,r_e}{r^2-r^2_e}
\ee
  and $r_e$ is the
horizon of the extremal black hole. One can easily check that this
solution indeed satisfies the conformal Klein-Gordon equation written
in extremal black hole background.

For the massless spinor we have
\be
C_{\al}(t,r,\phi)= e^{-iEt}e^{iL\phi}
(K_1 \,\psi_{1\alpha} + K_2 \,\psi_{2\alpha} ),
\ee
with
\[
\psi_{1}=\left(
\begin{array}{cc}
\mu M_{-Q,-\f14}(\kappa_e) \\
- (Q+\f14)\eta \kappa_e^{-\f{1}{2}} M_{-Q,\f14}(\kappa_e)\\
\end{array}\right)
\]
and
\[
\psi_{2}=\left(
\begin{array}{cc}
\mu M_{-Q,\f14}(\kappa_e) \\
-\eta \kappa_e^{-\f{1}{2}} M_{-Q,-\f14}(\kappa_e)\\
\end{array}\right),
\]
where $K_1, K_2$ are arbitrary constants.

\section{Symmetries of massless fields in BTZ black hole background}\label{Symmetry}

Any fixed vacuum solution \eqref{connect} of \eqref{zero} breaks
local HS symmetries to the global symmetries associated with the
stability subalgebra with the parameter $\gep_0(a,b|X)$ satisfying
\eqref{eps}. The
BTZ boundary condition \eqref{fact} restricts the space of solutions
of \eqref{eps} thus providing a
(non-local) mechanism of spontaneous symmetry breaking.
Namely, only those  symmetries remain well-defined upon
the factorization \eqref{fact} that commute to the Killing vector
$\xi_{\phi}$ responsible for the angle identification
\be\label{leftover}
[\xi(a,b), \xi_\phi]_{\star}=0\,,
\ee
where $\xi(a,b)$ is the generating parameter in \eqref{epss}. The
spaces of solutions of \eqref{leftover} are different for generic and
extremal black holes. We, therefore, consider these  cases
separately.

Let us start with the case of generic black hole $r_{+}^2-r_{-}^2>0$
and $\f{\p}{\p\phi}$ given by \eqref{phikil} with its star-product
realization $\xi_{\phi}$ \eqref{osckilf}. To make contact with
the gauge function \eqref{gWfunc}, we set $\al=1$ and $\gb=0$,
thus
\be
\xi_{\phi}=-\f12r_-\tau^{\gga\gd}L_{\gga\gd}-\f14r_+\tau^{\gga\gd}P_{\gga\gd}\,,\quad
\tau^{\gga\gd}=\left(
\begin{array}{cc}
0 & 1 \\
1 & 0 \\
\end{array}\right)\,.
\ee
To solve the equation \eqref{leftover} it is convenient to
introduce the new set of oscillators $p_{\al}$, $q_{\gb}$
\[
p_{1}=\f{1}{\sqrt{2}}(a_1+b_1)\,,\quad
p_2=\f{1}{\sqrt2}(a_1-b_1)\,,
\]
\be\label{newosc}
q_{1}=\f{1}{\sqrt{2}}(a_2-b_2)\,,\quad
q_2=\f{1}{\sqrt2}(a_2+b_2)\,
\ee
that satisfy the commutation relations
\be
[p_\al, p_{\gb}]_{\star}=[q^\al, q^{\gb}]_{\star}=0\,,\qquad
[p_{\al}, q^{\gb}]_{\star}=\delta_\al{}^\gb
\ee
and are chosen so that $\xi_\phi$ takes the following simple form
\be
\xi_{\phi}=-\f12A_{\al}{}^{\gb}p_{\gb}q^{\al}\,,\quad
A_{\al}{}^{\gb}=\left(
\begin{array}{cc}
r_++r_- & 0 \\
0 & r_--r_+ \\
\end{array}\right)\,.
\ee
Note, that since the oscillator commutation relations remain
unchanged, the same star-product formula \eqref{old} is valid
with $a$ and $b$ replaced by $p$ and $q$, respectively.

The equation \eqref{leftover} gives (cf \eqref{bilosc})
\be \label{genlo}
A_{\al}{}^{\gb}\Big( p_{\gb}\f{\p}{\p p_{\al}}+q^{\al}\f{\p}{\p
q^{\gb}}\Big)\xi(p,q)=0\,.
\ee
Infinitesimal HS symmetries we are interested in correspond to
local transformations with a finite number of space-time derivatives.
The corresponding symmetry generating parameters
$\xi(p,q)$ are described by  polynomial functions of the
oscillators. A class of  polynomial solutions of
\eqref{leftover} depends on the parameter
\be
\sigma=\f{r_{+}+r_{-}}{r_+-r_-}.
\ee
 There are  following
different cases:
\begin{itemize}
\item $\sigma \notin \mathbb{N}$
\end{itemize}
For any positive non-integer $\sigma$ the general solution of
\eqref{genlo} is
\be
\xi(p,q)=\sum R_{mn}(q_1p_2)^m(q_2p_1)^n\sim
\sum\tilde{R}_{mn}(\xi_{\phi})^{m}(\xi_t)^n\,,
\ee
where $R_{mn}$ are arbitrary constants. Note, that the conformal
algebra $sp(4)$, spanned by various bilinears of oscillators \eqref{newosc},
is broken to the $u(1)\oplus u(1)$ subalgebra spanned by the
 BTZ Killing vectors $\xi_{\phi}$ and $\xi_{t}$ (equivalently,
$q_1p_2$ and $q_2p_1$).
\begin{itemize}
\item $\sigma=2,3,\dots$
\end{itemize}
In the interesting case of positive integer $\sigma$
a larger class of  HS symmetries survives.
General solution of \eqref{genlo} is
\be
\xi(p,q)=\sum R_{n_1n_2m_1m_2}(q_1p_2)^{n_1}(q_2p_1)^{n_2}
(p_1p_2^\sigma)^{m_1}(q_1q_2^\sigma)^{m_2}\,.
\ee
The conformal algebra $sp(4)$ is still broken to $u(1)\oplus u(1)$.
The condition $\sigma=2,3,\dots$ imposes specific quantization
of the mass $M$ in terms of the
angular momentum $J$ since $\sigma=\sqrt{\f{M+J\gl}{M-J\gl}}$.
For this case it follows that
$\rho^{+}=(\rho^-)^\sigma$ which means that
one of the holonomy operators involved in the factorization
of BTZ black hole is the integral power of the other\footnote{We
are grateful to S. Carlip for drawing our attention to this fact.}.

\begin{itemize}
\item $\sigma=1$
\end{itemize}
This is the case of non-rotating black hole with $J=0$.
Polynomial solutions for $\xi(p,q)$ are
\be
\xi(p,q)=\sum R_{m_1m_2n_1n_2}
(q_1p_2)^{m_1}(q_2p_1)^{m_2}(p_1p_2)^{n_1}(q_1q_2)^{n_2}\,.
\ee
The distinguishing property of the non-rotating black hole is
that in this case a larger  part of the conformal symmetry survives.
It is generated by the bilinears $q_1p_2$, $q_2p_1$, $p_1p_2$, $q_1q_2$
and is isomorphic to $gl(2)$. In addition to BTZ Killing vectors, it has two
generators of special conformal transformations associated with
$b_1b_1$ and $b_2b_2$.

Let us proceed to the extremal case. The Killing vector of the
extremal black hole with $r_{-}=r_{+}=r_e$ responsible for the
angle identification is defined in \eqref{extrkill}. Using
\eqref{starkill} and setting $\al=1$, $\gb=0$, the expression for
$\xi_\phi$ in $p,q$ oscillators reads
\be
\xi_{\phi}=-r_ep_1q_2+\f12\left((p_1)^2-(q_1)^2\right).
\ee
Performing simple star-product calculations, we rewrite
\eqref{leftover} in the form
\be\label{extrsym}
r_e\Big(p^2\f{\p}{\p p^2}-q^1\f{\p}{\p q^1}\Big)
\xi-p^2\f{\p\xi}{\p q^1}+q^2\f{\p\xi}{\p p^1}=0\,.
\ee
The  cases with $r_e\neq 0$ and $r_e=0$ (i.e.,
$M=J=0$) require different consideration.
\begin{itemize}
\item $r_e\neq 0$
\end{itemize}
The general polynomial solution of \eqref{extrsym} is
\be
\xi(p,q)=\sum R_{mn}(p_1)^m(2r_eq_2-p_1)^m (q_1)^n\,.
\ee
One observes that, in addition to the usual $u(1)\oplus u(1)$ algebra
generated by Killing vectors $\xi_t$ and $\xi_\phi$, extremal black
hole has one Killing spinor generated by $q_1$. This is in accordance with
\cite{superBTZ} where supersymmetry of an extremal BTZ black hole
was found.
\begin{itemize}
\item $r_e=0$
\end{itemize}
The vacuum case of $M=J=0$ provides the black hole background with
the maximal number of supersymmetries and generic $\xi(p,q)$ of the form
\be
\xi(p,q)=\sum R_{mnk}(p_1)^m(q_1)^n(q_1q_2+p_1p_2)^k\,.
\ee
It has two exact supersymmetries \cite{superBTZ} generated by $p_1$
and $q_1$ and a part of conformal algebra spanned by $p_1p_1$,
$q_1q_1$, $q_1p_1$, $q_1q_2+p_1p_2$ which is isomorphic to $E_2
\oplus u(1)$, where $E_2$ is the algebra of motions of a two-dimensional
Euclidian plane.

Note that in our approach it is elementary to obtain
explicit formulae for the symmetry transformation laws.
The corresponding symmetry parameter \eqref{epss}
 for any generating parameter $\xi(a,b)$ results from the
 differentiation of the generating parameter \eqref{source}
 from Appendix A with
respect to the sources $\mu_{\al}$, $\eta_{\gb}$.

\section{Conclusion}

We have shown that the BTZ black hole can be concisely formulated
in terms of the star-product formalism underlying the present day
formulations of nonlinear HS gauge theories. Satisfying the
$o(2,1)\oplus o(2,1)$ zero-curvature condition, the BTZ black hole
is automatically an exact  solution of the nonlinear $3d$ HS gauge
theory. It is shown how the star-product formulation allows one to
solve free field equations in the black hole background.

The leftover higher spin and lower spin symmetries of massless fields
in the BTZ black hole background are found.
In the case of $M>0$ non-extremal BTZ black hole,
the conformal algebra $o(3,2)\sim sp(4)$ turns out to be
broken to the $u(1)\oplus u(1)$ subalgebra generated by BTZ Killing
vectors  and to $gl(2)$ in the cases of $J>0$ and
 $J=0$, respectively.
For
$\sigma=\sqrt{\f{M+J\gl}{M-J\gl}}=1,2,\dots$ the leftover
HS symmetries get enhanced. A physical interpretation
of this enhancement remains to be understood.
Our analysis of extremal BTZ black hole reproduces the
previously known lower spin (super)symmetries and determines
 their HS extensions.

We hope to
extend the obtained results in the following two, most likely
related, directions. Firstly, to Kerr solutions of nonlinear HS
gauge theories in four and higher dimensions and, secondly, to
BTZ-like solutions in the generalized space-times with matrix
coordinates which are $Sp(M)$ group manifolds in the $AdS$-like
case.

\section*{Acknowledgement}
We are grateful to S. Carlip, R. Metsaev and  I. Tyutin for useful
comments. This research was supported in part by INTAS Grants No
03-51-6346 and  05-7928, RFBR Grant No 05-02-17654, LSS No
4401.2006.2. V.D. and A.M. acknowledges financial support from
Landau Scholarship Foundation. V.D. and A.M. are grateful to
Dynasty Foundation for financial support, to ``Ettore Majorana''
Foundation and Center for Scientific Culture for kind hospitality
where a part of the work was done, and to Museo Storico della
Fisica e Centro Studi e Ricerche Enrico Fermi for Enrico Fermi
Junior award. M.V. is grateful to J. Azcarraga for hospitality at
the Physics Department of Valencia University, Spain, where a part
of this work was done.

\renewcommand{\theequation}{A.\arabic{equation}}
\renewcommand{\thesubsection}{A.}
\renewcommand{\thesection}{Appendix}

\section{}

\subsection{Action of (angular) momentum operator on
a scalar field}

To find the on-shell action of $L_{\al\gb}$ and $P_{\al\gb}$
generators on the scalar field let us consider the generating
parameter
\[
\xi=\exp(a_{\al}\mu^{\al}+b_{\al}\eta^{\al})
\]
with constant sources $\mu_{\al}$ and $\eta_{\al}$. As shown in
\cite{DV}, the global symmetry generators that result from
\eqref{epss} read as
\be\label{source}
\epsilon=\exp(a_\alpha\hat{\mu}^\alpha +
b_\alpha\hat{\eta}^\alpha)\,,
\ee
where
\begin{subequations} \label{g}
\begin{align}
\hat{\mu}_\alpha&=\frac12(W_1^{-1}+W_2)_\alpha{}^\beta\mu_\beta +
\frac12(W_1^{-1}-W_2)_\alpha{}^\beta\eta_\beta\,, \\
\hat{\eta}_\alpha&=\frac12(W_1^{-1}+W_2)_\alpha{}^\beta\eta_\beta +
\frac12(W_1^{-1}-W_2)_\alpha{}^\beta\mu_\beta\,.
\end{align}
\end{subequations}
The differentiation with respect to the sources
$\mu_{\al},\eta_{\al}$ gives the  generators of global $AdS_3$
symmetries
\begin{subequations} \label{adssym}
\begin{align}
\epsilon^P_{\alpha\beta}&=\left.\left(\frac{\partial^2}{\partial\mu^\alpha\partial\mu^\beta}
+\frac{\partial^2}{\partial\eta^\alpha\partial\eta^\beta}\right)\epsilon\right|_{\substack{\mu=0 \\ \eta=0}}, \\
\epsilon^L_{\alpha\beta}&=\left.\frac12\left(\frac{\partial^2}{\partial\mu^\alpha\partial\eta^\beta}
+\frac{\partial^2}{\partial\eta^\alpha\partial\mu^\beta}\right)\epsilon\right|_{\substack{\mu=0 \\ \eta=0}}.
\end{align}
\end{subequations}
Using \eqref{gaugeFock} and performing the star-products one
obtains
\be
\gd\Fock=\gep\star\Fock=\exp(b_{\al}\hat{\eta}^\alpha
+\f12\hat{\mu}_\alpha\hat{\eta}^\alpha)C(b+\hat{\mu}|X)\star\vac,
\ee
so that
\be\label{gaugeC}
\gd C(b|X)=C(b+\hat{\mu}|X)\exp(b_{\al}\hat{\eta}^\alpha
+\f12\hat{\mu}_\alpha\hat{\eta}^\alpha)\,.
\ee
As a result, from \eqref{adssym} and \eqref{gaugeC} one obtains the
following action of $AdS_3$ symmetries on the scalar field $C(X)$
\begin{subequations} \label{scalsym}
\begin{align}
\delta^P_{\alpha\beta}C(X)&=\left.\left(\frac{\partial^2}{\partial\mu^\alpha\partial\mu^\beta}
+\frac{\partial^2}{\partial\eta^\alpha\partial\eta^\beta}\right)
\left(C(\hat{\mu}|X)e^{\frac12\hat{\mu}_\gamma \hat{\eta}^\gamma}\right) \right|_{\substack{\mu=0 \\ \eta=0}}\,, \\
\delta^L_{\alpha\beta} C(X)&=\left.\frac12\left(\frac{\partial^2}{\partial\mu^\alpha\partial\eta^\beta}
+\frac{\partial^2}{\partial\eta^\alpha\partial\mu^\beta}\right)
\left(C(\hat{\mu}|X)e^{\frac12\hat{\mu}_\gamma \hat{\eta}^\gamma}\right) \right|_{\substack{\mu=0 \\
\eta=0}}\,
\end{align}
\end{subequations}
that gives
\begin{subequations}\label{scalact}
\begin{align}
\delta^P_{\alpha\beta}C(X)&=\frac12\left.\left(W_{1\,\alpha}{}^\gamma
W_{1\,\beta}{}^\delta + (W_2^{-1})_\alpha{}^\gamma
(W_2^{-1})_\beta{}^\delta\right)
\frac{\partial^2 C(\mu|X)}{\partial\mu^\gamma\partial\mu^\delta} \right|_{\mu=0}\,, \\
\delta^L_{\alpha\beta}C(X)&=\frac14\left.\left(W_{1\,\alpha}{}^\gamma
W_{1\,\beta}{}^\delta - (W_2^{-1})_\alpha{}^\gamma
(W_2^{-1})_\beta{}^\delta\right)
\frac{\partial^2 C(\mu|X)}{\partial\mu^\gamma\partial\mu^\delta}
\right|_{\mu=0}\,.
\end{align}
\end{subequations}
Using the equation of motion \eqref{Ceqtens}, for the scalar field
we have
\begin{equation}\label{scaleq}
dC(X)=\frac14 h^{\alpha\beta} \left.\frac{\partial^2
C(\mu|X)}{\partial\mu^\alpha\partial\mu^\beta} \right|_{\mu=0}
\end{equation}
or, equivalently,
\begin{equation}\label{scaleq2}
2h^{n}{}_{,\alpha\beta}\partial_{n}C(X)= \left.\frac{\partial^2
C(\mu|X)}{\partial\mu^\alpha\partial\mu^\beta} \right|_{\mu=0}.
\end{equation}
Using \eqref{gcalfi2} and taking \eqref{Sw1w2} into account,
the substitution of \eqref{scaleq2} into
\eqref{scalact} yields the following action of $AdS_3$ isometry
generators on the scalar field
\begin{subequations}
\begin{align}
\delta^P_{\alpha\beta}C(X) &= (\p_m S_{\alpha\gamma}
S_\beta{}^\gamma
- \p_m S_{\gamma\alpha} S^\gamma{}_\beta) g^{mn} \p_n C(X), \\
\delta^L_{\alpha\beta}C(X) &= \frac12(\p_m S_{\alpha\gamma}
S_\beta{}^\gamma + \p_m S_{\gamma\alpha} S^\gamma{}_\beta) g^{mn}
\p_n C(X).
\end{align}
\end{subequations}

\renewcommand{\theequation}{B.\arabic{equation}}
\renewcommand{\thesubsection}{B.}
\makeatletter \@addtoreset{equation}{subsection} \makeatother

\subsection{Star-product calculus}

Here we collect some useful  star-product formulae used throughout this paper
\begin{align} \label{bilosc}
a_{\alpha}a_{\beta} \star f(a,b)&= a_{\alpha}a_{\beta} f(a,b) +
\frac12 \left( a_{\alpha}\frac{\partial}{\partial b^{\beta}} +
a_{\beta}\frac{\partial}{\partial b^{\alpha}} \right) f(a,b)
+ \frac14 \frac{\partial^2}{\partial b^{\alpha}\partial b^{\beta}} f(a,b), \notag \\
b_{\alpha}b_{\beta} \star f(a,b)&= b_{\alpha}b_{\beta}
f(a,b) + \frac12 \left( b_{\alpha}\frac{\partial}{\partial
a^{\beta}} + b_{\beta}\frac{\partial}{\partial a^{\alpha}} \right)
f(a,b)
+ \frac14 \frac{\partial^2}{\partial a^{\alpha}\partial a^{\beta}} f(a,b), \\
a_{\alpha}b_{\beta} \star f(a,b)&= a_{\alpha}b_{\beta}
f(a,b) + \frac12 \left( a_{\alpha}\frac{\partial}{\partial
a^{\beta}} + b_{\beta}\frac{\partial}{\partial b^{\alpha}} \right)
f(a,b) + \frac14 \frac{\partial^2}{\partial a^{\beta}\partial
b^{\alpha}} f(a,b).\notag
\end{align}
Let us now calculate the generating function \eqref{CbBTZ}.
According to the prescription given in Section \ref{Fockmod}
\be\label{1}
\Fock=g^{-1}(a,b|HW_1, W_2)\star C(b)\star\vac\Big|_{\substack{\al=1 \\ \gb=0}}\,,
\ee
where $C(b)$ is defined in \eqref{Cb}. The direct calculations are
quite cumbersome. In practice, it is convenient to eliminate Lorentz
term in the gauge function $g(a,b|HW_1, W_2)$. With the aid of
\eqref{lorg}, \eqref{gom} one can rewrite \eqref{1} in the form
\be \label{decompg}
\Fock = \Lambda^{-1}(a,b|V)\star g^{-1}(a,b|HW_1V, V^{-1}W_2)\star
C(b)\star\vac\Big|_{\substack{\al=1 \\ \gb=0}}\,.
\ee
Let us choose Lorentz transformation matrix $V_{\gamma\delta}$ such that
\[
HW_1V=V^{-1}W_2=\sqrt{S}.
\]
Then the gauge function takes the form
\be \label{g0func}
g_{0}(a,b|S)=g(a,b|\sqrt{S},
\sqrt{S})=\f{4}{\det||\sqrt{S}+1)||}\exp\Big(
-\Pi^{\al\gb}(\sqrt{S})(a_\al a_{\gb}+b_\al b_\gb)\Big)\,,
\ee
where $S_{\gamma\delta}$ is given in \eqref{newSm}.
Thus using \eqref{lorFock}, the generating function can be
calculated as
\be\label{Clor}
|C(b_\gamma|X)\rangle=|C_0((V^{-1}_{0})_\gamma{}^{\gd}b_{\gd}|X)\rangle\,,
\ee
where $V_0$ is the Lorentz transformation matrix at $\al=1$,
$\gb=0$ which has the following form
\be\label{V}
V_{0\,\gga\gd}=-\sqrt{\f{(u+x)(y-v)}{{2(u+1)}}}\left(
\begin{array}{cc}
\mu(r) &  \mu(r)\f{x-u-1}{y-v} \\
\eta(r)(u-x)& \eta(r)\f{(u-x)(u+x+1)}{y-v} \\
\end{array}\right)
\ee
and
\[
|C_0(b|X)\rangle=g_{0}^{-1}(a,b|S)\star C(b)\star\vac\Big|_{\substack{\al=1 \\ \gb=0}}\,.
\]
To evaluate star-products of  Gaussian exponentials of the form
\be
F(b)\star |0\rangle\langle 0| = \sqrt{\det||1-f^2||}
e^{f^{\alpha\beta}(a_\alpha a_\beta + b_\alpha b_\beta)} \star
e^{m_{\gamma\delta}b^\gamma b^\delta+n_\gamma b^\gamma} \star
|0\rangle\langle 0|
\ee
one uses \eqref{old} to obtain  by  simple Gaussian integration
\begin{align}\label{genstar}
F(b) &= \sqrt{\frac{\det||1-f^2||}{\det(A)}} \exp\left(
\left[(f^{-1}+2m)\frac{1}{f+f^{-1}+4m}(f+2m) - m \right]_{\alpha\beta} b^{\alpha}b^{\beta} \right.\notag \\
&+ \left[ \frac{1}{f+f^{-1}+4m}(f^{-1}-f) \right]_{\alpha\beta}
n^{\alpha}b^{\beta} +\left. \left[ \frac{1}{f+f^{-1}+4m}
\right]_{\alpha\beta} n^\alpha n^\beta \right),
\end{align}
where
$$
(f^{-1})_\alpha{}^\beta f_{\beta}{}^{\gamma} =
\delta_{\alpha}{}^{\gamma}
$$
and
$$
A_{\alpha\beta}=\varepsilon_{\alpha\beta} + f_\alpha{}^\gamma
f_{\gamma\beta} + 4f_\alpha{}^\gamma m_{\gamma\beta}.
$$
Using \eqref{Cb}, \eqref{g0func} and \eqref{genstar} we obtain the following
result
\begin{align} \label{CbX1}
F(b)&=\left(\f{\gb}{y-v}\right)^{\f12}\int\limits_{-\infty}^{\infty}ds\int
dw\,
s^{2P-\f12}w^{2Q-\f12}e^{-\f{u+x}{4(y-v)}s^2-\gb^2\f{u-x}{y-v}w^2+\f{\gb}{y-v}
sw} \notag \\
&\times\exp\left(B_{\gga\gd}\,b^{\gga}b^{\gd} +
\f{A_{\gga}b^{\gga}}{(y-v)\sqrt{2(\al u-\gb y+1)}}\right)
\end{align}
with
$$
A_{\gga} = \left(
                  \begin{array}{c}
                    2\gb(\al(y-v)+\gb(x-u))w+(\gb+y-v)s \\
                    2\gb(\al(x-u)+\gb(y-v)-1)w+(u+x+\al)s \\
                  \end{array}
                \right)
$$
and
\[
B_{\gga\gd}=\f{\gb}{y-v}m_{\gga\gd} +\f{y-v+\beta}{2(y-v)(\al
u-\gb y+1)}S_{(\gga\gd)}\,,
\]
where parentheses denote index symmetrization.

By redefining the integration variable $\gb w\to w$ (reabsorbing
a $\gb$-dependent factor into an integration constant) and setting
then $\al=1$, $\gb=0$ we obtain
\begin{align}\label{CbX2}
C_0(b|X) &=(y-v)^{-\f12}\int\limits_{-\infty}^{\infty}ds\int dw \,
s^{2P-\f12}w^{2Q-\f12}e^{-\f{u+x}{4(y-v)}s^2-\f{u-x}{y-v}w^2+\f{1}{y-v}sw}
\cdot
e^{\hat{B}_{\gamma\delta}\,b^{\gamma}b^{\delta}} \notag  \\
& \times\exp\left(\f{b^1(2w+s)(y-v)+ b^2
(s(u+x+1)+2w(x-u-1))}{(y-v)\sqrt{2(u+1)}}\right),
\end{align}
with
\[
\hat{B}_{\gamma\delta}=\f12\left(
\begin{array}{cc}
\f{y-v}{u+1} & \f{x}{u+1} \\
\f{x}{u+1} & -\f{y+v}{u+1}-\f{2}{y-v} \\
\end{array}\right).
\]
Performing Lorentz transformation \eqref{Clor} and using BTZ
coordinates \eqref{btzcoord} we finally obtain \eqref{CbBTZ}.

Let us also note, that a convenient parametrization of a
$Sp(2)$-valued matrix $S_{\alpha\beta}$ is
\[
S_{\alpha\beta}=\cosh(p)\varepsilon_{\alpha\beta}+\sinh(p)\kappa_{\alpha\beta}\,,
\]
where $\gk_{\al\gb}=\gk_{\gb\al}$ and
$\f12\gk_{\al\gb}\gk^{\al\gb}=-1$. One  can see that the $n^{th}$
power of $S$ is
\[
(S^n)_{\al\gb}=\cosh(np)\gep_{\al\gb}+\sinh (np)\gk_{\al\gb}\,.
\]
As a result,
\[
(\sqrt{S})_{\al\gb}=\cosh\left(\frac{p}{2}\right)\varepsilon_{\alpha\beta}
+\sinh\left(\frac{p}{2}\right)\kappa_{\alpha\beta}\,.
\]
Also one finds that the matrix $\Pi=\frac{\sqrt{S}-1}{\sqrt{S}+1}$
is
\[
\Pi_{\alpha\beta}=\tanh\left(\frac{p}{4}\right)\kappa_{\alpha\beta}\,.
\]


\begin{thebibliography}{28}

\bibitem{1963}
A. Staruszkiewicz, \textit{Acta Phys. Polon.} \textbf{24} (1963) 734

\bibitem{1966}
H. Leutwyler, \textit{Nuovo Cim.} {\bf 42} (1966) 159

\bibitem{3d1}
S. Deser, R. Jackiw and G. 't Hooft, \textit{Annals Phys.} {\bf 152}
(1984) 220

\bibitem{3d2}
S. Deser and R. Jackiw, \textit{Annals Phys.} {\bf 153} (1984) 405;
\textit{Commun.Math.Phys.} {\bf 118} (1988) 495

\bibitem{3d3}
G. 't Hooft, \textit{Commun.Math.Phys.} {\bf 117} (1988) 685

\bibitem{Townsend}
A. Ach\'{u}carro and P. K. Townsend, \textit{Phys.Lett.} {\bf B180}
(1986) 89

\bibitem{Wit}
E. Witten, \textit{Nucl.Phys.} {\bf B311} (1988) 46;
\textit{Nucl. Phys.} \textbf{B323} (1989) 113;
\textit{Commun.Math.Phys.} {\bf 137} (1991) 29

\bibitem{Carlip}
S. Carlip, \textit{J.Korean Phys.Soc.} {\bf 28} (1995) S447-S467,
gr-qc/9503024

\bibitem{BTZ}
M. Banados, C. Teitelboim and J. Zanelli, \textit{Phys.Rev.Lett.}
{\bf 69} (1992) 1849, hep-th/9204099

\bibitem{nobh}
D. Ida, \textit{Phys.Rev.Lett.} {\bf 85} (2000) 3758, gr-qc/0005129

\bibitem{BHTZ}
M. Banados, M.Henneaux, C. Teitelboim and J. Zanelli,
\textit{Phys.Rev.} {\bf D48} (1993) 1506, gr-qc/9302012

\bibitem{3HS}
M.A. Vasiliev, \textit{Mod.Phys.Lett.} {\bf A7} (1992) 3689-3702

\bibitem{PV}
S.F. Prokushkin and M.A. Vasiliev,
\textit{Nucl.Phys.} {\bf B545} (1999) 385, hep-th/9806236

\bibitem{Gol}
M.A. Vasiliev, \emph{Higher Spin Gauge Theories: Star Product and
AdS Space}, Contributed article to Golfand's Memorial Volume ``Many
faces of the superworld'', ed. by M. Shifman, World Scientific
Publishing Co Pte Ltd, Singapore, 2000, hep-th/9910096.

\bibitem{SS}
E. Sezgin and P. Sundell, \emph{An Exact Solution of 4D Higher-Spin
Gauge Theory}, hep-th/0508158

\bibitem{F}
C. Fronsdal, Preprint UCLA/85/TEP/10, in Essays on Supersymmetry,
Reidel, 1986 (Mathematical Physics Studies, v.8)

\bibitem{BLS}
I. Bandos, J. Lukierski and D. Sorokin, \textit{Phys.Rev.} {\bf D61} (2000) 045002,
hep-th/9904109

\bibitem{BHS}
M.A. Vasiliev, \textit{Phys.Rev.} {\bf D66} (2002) 066006, hep-th/0106149

\bibitem{BLPS}
I. Bandos, J. Lukierski, C. Preitschopf and D. Sorokin, \textit{Phys.Rev.}
{\bf D61} (2000) 065009, hep-th/9907113

\bibitem{DV}
V.E. Didenko and M.A. Vasiliev, \textit{J.Math.Phys.} {\bf 45} (2004) 197-215,
hep-th/0301054

\bibitem{PST}
M. Plyushchay, D. Sorokin and M. Tsulaia, \textit{JHEP} {\bf 0304} (2003) 013,
hep-th/0301067

\bibitem{SV}
O.V. Shaynkman and M.A. Vasiliev, \textit{Theor.Math.Phys.} {\bf 128}
(2001) 1155-1168; (\textit{Teor.Mat.Fiz.} {\bf 128} (2001) 378-394), hep-th/0103208

\bibitem{scalarBTZ1}
K. Ghoroku and A. L. Larsen, \textit{Phys.Lett.} {\bf B328} (1994)
28-35, hep-th/9403008

\bibitem{scalarBTZ2}
I. Ichinose and Y. Satoh, \textit{Nucl.Phys.} {\bf B447} (1995)
340, hep-th/9412144

\bibitem{BTZspinor}
S. Das and A. Dasgupta, \textit{JHEP} {\bf 9910} (1999) 025,
hep-th/9907116

\bibitem{BSS}
D. Birmingham, I. Sachs and S.N. Solodukhin,
\textit{Phys.Rev.Lett.} {\bf 88} (2002) 151301, hep-th/0112055

\bibitem{Bars}
I.Bars and M.G\"{u}naydin, \textit{Commun.Math.Phys.} {\bf 91} (1983) 31

\bibitem{V91}
M.A. Vasiliev, \textit{Class.Quant.Grav.} {\bf 11} (1994) 649-664

\bibitem{ryzhik}
I.S. Gradshtein and I.M. Ryzhik, \emph{Tables of integrals, sums,
series and products}, Fiz.Mat.Lit., Moscow, 1963.

\bibitem{extr1}
J. Gamboa and F. M\'{e}ndez, \textit{Class.Quant.Grav.} {\bf 18}
(2001) 225-232, hep-th/0006020

\bibitem{extr2}
S. Lepe, F. M\'{e}ndez, J. Saavedra and L. Vergara, \textit{Class.Quant.Grav.}
 {\bf 20} (2003) 2417-2428, hep-th/0302035

\bibitem{troost}
J. Troost, \textit{JHEP} {\bf 0209} (2002) 041, hep-th/0206118

\bibitem{Vilenkin}
N.A Vilenkin, \emph{Special functions and the theory of group
representations}, Providence, American Mathematical Society, 1968

\bibitem{superBTZ}
O. Coussaert and M. Henneaux, \textit{Phys.Rev.Lett.} {\bf 72}
(1994) 183-186, hep-th/9310194

\end{thebibliography}
\end{document}